%% file: bigi2.tex
\documentstyle[preprint,aps,eqsecnum]{revtex}

\begin{document}
\input epsf.sty
\input sanda.tex
\begin{titlepage}
\renewcommand{\thefootnote}{\fnsymbol{footnote}}

\begin{flushright}
UND-HEP-98-BIG\hspace*{.2em}05\\
EPNU-98-41\\
\end{flushright}
\vspace{.3cm}
\begin{center} \Large  
{\bf Comments On The Sign of \cp~Asymmetries}
\end{center}
\vspace*{.3cm}
\begin{center} {\Large
I. I. Bigi $^{a}$, A. I. Sanda $^{b}$}\\
\vspace{.4cm}
{\normalsize 
$^a${\it Physics Dept.,
Univ. of Notre Dame du
Lac, Notre Dame, IN 46556, U.S.A.}\\
$^b$ {\it  Physics Dept., Nagoya University, 
Nagoya 464-01,Japan}}
\\
\vspace{.3cm}
e-mail addresses:\\
{\it bigi@undhep.hep.nd.edu, 
sanda@eken.phys.nagoya-u.ac.jp} 

\vspace*{.4cm}
{\Large{\bf Abstract}}\\
\end{center} 
Several experiments are about to measure the \cp~ asymmetry
for $B\to\psi K_S$ and other decays. The standard model together 
with the Kobayashi-Maskawa ansatz for \cp~ violation predicts the 
sign as well as the magnitude for this asymmetry. In this note we 
elucidate the physics and conventions which lead to the prediction 
for the sign of the asymmetry.

\vspace*{.2cm}
\vfill
\noindent
\end{titlepage}

\newpage

%
%%%%%%%%%%%%%% 
\section{The Problem}
%%%%%%%%%%%%

It has been pointed out some time ago \cite{CARTER} 
that within the KM ansatz large \cp~asymmetries have to arise 
in B decays involving $B^0 - \bar B^0$ oscillations. 
In particular the channel $B_d \to \psi K_S$ combines 
a striking experimental signature with a clean theoretical 
interpretation \cite{BS}; therefore it is often referred to 
as the "golden mode".

A first experimental information on the \cp~asymmetry 
in it has recently 
become available\cite{opal,cdfl}. The asymmetry is defined as 
follows 
\be 
A_{\psi K_S}=\frac{ \Gamma(\overline B_d(t)\to\psi K_S)-
\Gamma(B_d(t)\to\psi K_S)}
{\Gamma(\overline B_d(t)\to\psi K_S)+\Gamma(B_d(t)\to\psi K_S)}
=\Im\left( \frac{q}{p}
\overline\rho(\psi K_S)\right) \sin (\Delta M_{B_d} t) 
\mlab{10.26}
\ee
where $\overline\rho(\psi K_S)$ denotes the ratio 
of transition amplitudes 
\be
\overline\rho(\psi K_S) =\frac{A(\overline B\to\psi K_S)}
{A(B\to\psi K_S)}.
\ee 
while $q$ and $p$ relate the mass eigenstates to the flavour eigenstates $B_d$
and $\overline B_d$ (see below).
Two groups found 
\be
\Im\left( \frac{q}{p}
\overline\rho(\psi K_S)\right)=\left\{
{+({3.2{+1.8\atop -2.0}\pm 0.5}) {\rm ~~OPAL\; 
Collaboration\cite{opal}}
\atop
{+(1.8\pm 1.1\pm0.3)}{\rm ~~CDF\; 
Collaboration\cite{cdfl}}}\right.
\ee
In reality the observable $\Im\left( \frac{q}{p}
\overline\rho(\psi K_S)\right)$ is bounded by
-1 and +1.  It falls outside this range in the data 
since they 
require subtracting a background of uncertain size. Thus the 
numbers have to be taken with a grain of salt. Even so they 
indicate that values close to +1 are very strongly disfavoured. 

This raises the following question: 
Can one 
{\em predict} also 
the {\em sign} in addition to the size of the asymmetry? 

At first one might think that to be impossible. For 
$A_{\psi K_S}$ is a product of 
$\Im\left( \frac{q}{p}
\overline\rho(\psi K_S)\right)$ and 
sin$(\Delta M_{B_d} t)$, see \mref{10.26}. The observable sign of 
$A_{\psi K_S}$ thus depends on the signs of both 
$\Im\left( \frac{q}{p}
\overline\rho(\psi K_S)\right)$ and $\Delta M_{B_d}$, 
and the sign of the latter cannot be defined nor 
determined {\em experimentally} in a feasible way -- 
in contrast to the case with kaons. 

In this note we will show that the overall sign of the 
asymmetry $A_{\psi K_S}$ -- and likewise for 
other asymmetries -- can be predicted within a given theory 
for $\Delta B=2$ dynamics. 
For those who have thought about this question carefully,
this has been known for a long time. 
Indeed it has been discussed in \cite{GKN}. Yet it has not been explained
with all its aspects and in full detail.
 As the question becomes experimentally relevant,
we feel that it is important to display all the subtleties involved.
The paper will be organized as follows: in 
Sect. \ref{DELTAM} we discuss the theoretical evaluation of 
$\Delta M_K$, $\Delta M_{B_d}$, and $\Delta M_{B_s}$ 
together with $\Delta\Gamma_{B_S}$; in Sect. \ref{PHASE}    
we analyze the phase of $(q/p)\bar \rho$;
in Sect. \ref{OT} we address other conventions or proposals;
in Sect. \ref{TRIANGLES} we lay out a proper definition of the angles in the unitarity triangle before giving a summary in Sect. \ref{SUMMARY}.

%%%%%%%%%%%%%%
\section{The sign of $\Delta M$ \label{DELTAM}}
%%%%%%%%%%%

For our discussion to be more transparent, we adopt a 
formalism and conventions which are applicable equally 
to the $K^0 - \bar K^0$ and the $B^0 - \bar B^0$ complexes. 

Consider a neutral meson $P$ carrying a quantum number
$F=-1$; it can denote a $K^0 $ or 
$B^0$. The time evolution of a state being a 
mixture of $P$ and $\bar P$ is given by 
\be 
i \hbar \frac{\partial}{\partial t} \Psi(t) = {\cal H}\Psi(t) 
\mlab{Schroed2} 
\ee 
where $\Psi(t)$ is restricted to the subspace of $P$ and 
$\overline P$:
\be 
\Psi (t) = \left( \matrix {a(t) \cr \bar b(t) \cr } 
\right). 
\ee
Assuming \cpt~ symmetry, the matrix ${\cal H}$ is given by
\be 
{\cal H}= {\bf M} - \frac{i}{2} {\bf \Gamma} =
\left( \matrix {M_{11} - \frac{i}{2}\Gamma _{11} & 
M_{12} - \frac{i}{2}\Gamma _{12} \cr 
M_{12}^* - \frac{i}{2}\Gamma _{12}^* & 
M_{11} - \frac{i}{2}\Gamma _{11} \cr } \right), 
\mlab{CPTMass} 
\ee 
where
\ba 
M_{11} &=& M_P + 
\sum _n {\cal P}\left[ \frac{|\langle n;out|
H_{weak}|P\rangle |^2}
{M_P - M_n}\right] \nn
M_{12}&=& \matel{P}{H_{SW}}{\overline P} + 
\sum _n {\cal P}
\left[ \frac{\matel{P}{H_{\Delta F=1}}{n;out}
\matel{n;out}{H_{\Delta F=1}}{\overline P}} 
{M_P - M_n}\right] \nn
\Gamma _{11} &=& 2\pi \sum _n\delta(M_P-
M_n)|\matel{n;out}{H_{\Delta F=1}}{P}|^2 \nn
\Gamma _{22} &=& 2\pi \sum _n\delta(M_P-
M_n)|\matel{n;out}{H_{\Delta F=1}}{\overline P}|^2 \nn
\Gamma _{12}&=&2\pi
\sum _n\delta(M_P-M_n)\matel{P}{H_{\Delta F=1}}{n;out} 
\matel{n;out}{H_{\Delta F=1}}{\overline P}.
\mlab{6.8}
\ea
The coupled Schr\" odinger equations 
\mref{Schroed2} are best solved by diagonalizing 
the matrix ${\cal H}$. We find that 
\ba 
|P_1\rangle &= & 
p |P\rangle + q |\overline P\rangle , \nn   
|P_2\rangle &= &
p |P\rangle - q |\overline P\rangle ,
\mlab{P1P2gen} 
\ea   
are mass eigenstates with eigenvalues
\ba 
M_1 - \frac{i}{2}\Gamma _1 &=& 
M_{11}-\frac{i}{2}\Gamma_{11}+
\frac{q}{p}(M_{12}-\frac{i}{2}\Gamma_{12})\nn
M_2 - \frac{i}{2}\Gamma _2 &=& 
M_{11}-\frac{i}{2}\Gamma_{11}-
\frac{q}{p}(M_{12}-\frac{i}{2}\Gamma_{12})\nn
\mlab{EV} 
\ea
as long as 
\be
\left(\frac{q}{p}\right)^2=\frac{M_{12}^*-\frac{i}{2}\Gamma_{12}^*}
{M_{12}-\frac{i}{2}\Gamma_{12}}
\mlab{qoverp}
\ee
holds. Obviously, there are two solutions to this condition, namely
\be
\frac{q}{p} = \pm \sqrt{\frac{M_{12}^*-\frac{i}{2}\Gamma_{12}^*}
{M_{12}-\frac{i}{2}\Gamma_{12}}}.
\mlab{pmsign}
\ee

Choosing the negative rather than the positive sign in \mref{pmsign}
is equivalent to interchanging the labels $1\leftrightarrow 2$ of the mass eigenstates, see \mref{P1P2gen}, \mref{EV}. 

This binary ambiguity is a special case of a more general one. For 
antiparticles are defined only up to a phase; adopting a different phase 
convention - \eg going from $\cp|P\ket=|\overline P\ket$ to
$\cp|P\ket=e^{i\xi}|\overline P\ket$ - will modify 
$M_{12}-\frac{i}{2}\Gamma_{12}\equiv \mat{P}{{\cal H}}{\overline P}$:
\be 
M_{12} - \frac{i}{2}\Gamma_{12} \; \; \to \; \; 
e^{i\xi}[ M_{12} - \frac{i}{2}\Gamma_{12}] 
\ee   
and thus
\be
\frac{q}{p}\to e^{-i\xi}\frac{q}{p},
\ee
yet leave the combination
$\frac{q}{p}(M_{12} - \frac{i}{2}\Gamma _{12})$
invariant. This is as it should be since 
the differences in mass and width
\ba
M_2-M_1&=&-2\Re\left(\frac{q}{p}(M_{12} - \frac{i}{2}\Gamma _{12})\right)\nn
\Gamma_2-\Gamma_1&=&4\Im\left(\frac{q}{p}(M_{12} - \frac{i}{2}\Gamma _{12})\right)
\mlab{ei23}
\ea
being observables, have to be insensitive to the arbitrary phase of $\overline P$.

Some comments are in order to elucidate the situation:
\begin{itemize}
\item We can define the labels 1 and 2 such that 
\be 
\Delta M \equiv M_2-M_1 > 0
\ee
is satisfied. Once this {\em convention} has been adopted, it becomes a 
sensible question whether
\be
\Gamma_2>\Gamma_1~~~~~~~~~~~~~~~~~~~~or~~~~~~~~~~~~~~~\Gamma_2<\Gamma_1
\ee
holds, \ie whether the heavier state is shorter or longer lived.
\item
In the limit of \cp~ invariance the two mass eigenstates are \cp~eigenstates as well, and we can raise another meaningful question: is the heavier state \cp~
even or odd?
%B1
Since \cp~invariance implies $\arg\frac{M_{12}}{\Gamma_{12}} = 0$, 
$\frac{q}{p}$ becomes a pure phase: $|\frac{q}{p}| =1$. 
It is then convenient to adopt a phase convention s.t. 
$M_{12}$ is real; it leads to $\frac{q}{p}=\pm1$ and  
$\cp|P\ket= \pm|\overline P\ket$ as remaining choices.
\begin{itemize}
\item
With $\frac{q}{p}=+1$ we have
\ba 
|P_1\rangle &= & 
\frac{1}{\sqrt{2}}(|P\rangle +  |\overline P\rangle)  \nn   
|P_2\rangle &= &
\frac{1}{\sqrt{2}}(|P\rangle -  |\overline P\rangle) 
\mlab{B1} 
\ea   
For $\cp|P\ket= |\overline P\ket$, $P_1$ and $P_2$ are \cp~even and odd, 
respectively and therefore
\ba
M_--M_+=M_2-M_1&=&-2\Re \frac{q}{p}(M_{12} - \frac{i}{2}\Gamma _{12})
\nn
&=&-2M_{12}
\mlab{B2}
\ea
For $\cp|P\ket = -|\overline P\ket$, on the other hand,
$P_1$ and $P_2$ switch roles; 
i.e. $P_1$ and $P_2$ are \cp~odd and even now. Thus 
\be
M_--M_+=M_2-M_1= 2M_{12}
\mlab{B3}
\ee
\item
Alternatively we can set $\frac{q}{p}= -1$:
\ba 
|P_1\rangle &= & 
\frac{1}{\sqrt{2}}(|P\rangle - |\overline P\rangle)  \nn   
|P_2\rangle &= &
\frac{1}{\sqrt{2}}(|P\rangle + |\overline P\rangle) 
\mlab{B4} 
\ea   
while maintaining $\cp|P\ket = |\overline P\ket$; $P_1$ and $P_2$ 
are then \cp~odd and even, respectively. Accordingly
\ba
M_--M_+=M_1-M_2&=& 2\Re \frac{q}{p}(M_{12} - \frac{i}{2}\Gamma _{12})
\nn
&=&-2M_{12}
\mlab{B5}
\ea
\item
\mref{B2} and \mref{B5} on one side and \mref{B3} on the other 
do not coincide on the surface; yet,
we will see below  that the theoretical prediction for $M_{12}$ changes 
sign depending on the choice of $\cp|P\ket=\pm|\overline P\ket$.
Thus they all agree, of course.
\end{itemize}

\item
Within a given theory for the $P-\overline P$ complex, 
we can evaluate $M_{12}$, as discussed below. 
Yet some care has to be applied in 
interpreting such a result. For expressing mass eigenstates explicitely in 
terms of flavour eigenstates involves some conventions; see the examples above. Once we adopt a certain convention, we have to stick with it; yet our original choice cannot influence observables. It is instructive to trace how this comes about.
\end{itemize}

The {\em relative} phase between $M_{12}$ and $\Gamma_{12}$ on the other hand 
represents an observable quantity describing indirect \cp~violation. 
Therefore, we adopt the notation
\be 
M_{12} = \overline M_{12} e^{i\xi},~~~
\Gamma_{12}=\overline \Gamma_{12}e^{i\xi}e^{i\zeta},~~~\mbox{and}~~~
\frac{\Gamma_{12}}{M_{12}}=\frac{\overline \Gamma_{12}}{\overline M_{12}}
e^{i\zeta}=re^{i\zeta}
\mlab{219}
\ee
The sign of $\overline M_{12} $ and $\overline \Gamma_{12}$ are fixed 
such that $\xi$, and $\xi+\zeta$ are
restricted to lie between
$-\frac{\pi}{2}$ and $\frac{\pi}{2}$, respectively; \ie, the real quantities
$\overline M_{12}$ and $\overline \Gamma_{12}$ are a priori allowed to be 
{\em negative} as well as {\em positive}! A relative minus sign between
$M_{12}$ and $\Gamma_{12}$ is of course physically significant,
while the absolute sign is not. Yet, we will see that the absolute sign provides us with a useful bookkeeping device.

%%%%%%%%%%%%%%
\subsection{$\Delta M_K$}
%%%%%%%%%%%%
The two kaon mass eigenstates can unambiguously be labelled 
by their 
lifetimes as $K_L$ and $K_S$. One can then address the 
question how they differ in other properties: data reveal 
that (i) the $K_L$ is ever so slightly heavier: $M_L > M_S$ 
and (ii) the $K_S$ [$K_L$] is mainly \cp~even [odd]. We then 
adopt the conventions 
\be
\Delta M=M_2-M_1,~~~~~~~\mbox{ and }~~~~~~
\Delta\Gamma=\Gamma_1-\Gamma_2
\ee
which make both differences positive: 
\be
\Delta M_K=M_L-M_S>0,~~~~~\Delta\Gamma_K=\Gamma_S-\Gamma_L>0.
\ee
Since \cp~violation is very small in the K system -
$\zeta_K=\arg(\Gamma_{12}^K/M_{12}^K)\ll 1$ - we deduce from
\mref{ei23} in the notation of \mref{219}:
\be
\Delta M_K \simeq -2\overline M^K_{12}~~~~~\mbox{and}~~~~~
\Delta \Gamma_K \simeq 2\overline \Gamma^K_{12}.
\ee
We will show below that the standard model box diagram 
indeed reproduces the `correct', i.e. observed 
sign for $\Delta M_K$ (and -- more surprisingly -- even 
for $\Delta \Gamma _K$).

%%%%%%%%%%%%%%
\subsection{$\Delta M_{B_d}$}
%%%%%%%%%%%
The situation is qualitatively different here. 
While the two mass eigenstates will have different 
lifetimes, no useful experimental information 
exists on that. It is actually quite unlikely that such 
information will become available in the foreseeable 
future, since $\Delta \Gamma_{B_d}/\Gamma_{B_d}$ is 
estimated not to exceed the 1\% level 
\footnote{Remember that $\Gamma _S \gg 
\Gamma _L$ represents a kinematical accident!}. 
Thus we have to rely on a theoretical prediction of 
$\Delta M_{B_d}$.  

We can confidently predict for ${B_d}$ mesons 
\be 
|r_{B_d}|  \ll 1.
\ee  
Since we have to leading nontrivial order in $r_{B_d}$ 
\be 
\left. \frac{q}{p}\right| _{B_d}\simeq
\sqrt{\frac{(M_{12}^{B_d})^*}{M^{B_d}_{12}}}(1-\frac{r_{B_d}}{2}
\sin\zeta _{B_d}),
\ee
we get from \mref{ei23}
\be
\Delta M_{B_d}=-2\overline M^{B_d}_{12}~~~~~\mbox{and}~~~~~
\Delta \Gamma_{B_d}=2\overline \Gamma_{12}^{B_d} \cos\zeta_{B_d}.
\ee
Note that the expression for $\Delta M_{B_d}$ 
is similar to the kaon case, albeit for a different reason, 
namely $|\Gamma _{12}/M_{12}| \ll 1$ rather than 
$\arg\frac{\Gamma_{12}}{M_{12}} \ll 1$;
the latter quantity is actually not small for $B_d$ mesons. 

\subsection{$\Delta M_{B_s}$ and $\Delta \Gamma_{B_s}$}
There are two features that distinguish $B_s$ from $B_d$ decays in ways that are quite significant for our present discussion.
\begin{itemize}
\item
While $r_{B_s}\ll 1$ holds also for ${B_s}$ mesons, $\Delta\Gamma/\Gamma$ might not be that small. It has been estimated
\cite{URALTSEV}
\be
\frac{\Delta\Gamma}{\hat\Gamma}|_{B_s}\sim 0.18\left(\frac{f_{B_s}}{200MeV}\right)^2,~~~~\hat\Gamma=\half(\Gamma_{B_{s,1}}+\Gamma_{B_{s,2}}).
\ee
Such a difference in the lifetimes of the two $B_s$ mass eigenstates might actually become observable! This would raise the question whether the heavier state is longer or shorter lived.
\item
The $\Delta B=2$ effective operator for $B_s$ obtained from the box diagram is dominated by the quarks of the second and third families only. It thus predominantly conserves \cp~invariance making the two $B_s$ mass eigenstates approximately \cp~eigenstates as well. This has two consequences:
\begin{enumerate}
\item
Rather than fit the decay rate evolution for $B_s\to D_s^{(*)}\pi$ or 
$B_s\to l\nu D_s^{(*)}$ with two separate exponentials, we can compare the lifetimes of the \cp~even eigenstate, as measured in $B_s\to\psi\eta$, with the average lifetime obtained from $B_s\to l\nu D_s^{(*)}$. We can also compare the lifetimes in $B_s\to \psi\phi$ for S- and P- wave final states being \cp~even and odd, respectively.
\item
We can raise the issue whether the \cp~even state is longer or shorter lived.
\end{enumerate}
\end{itemize}

As for $B_d$ mesons, we have 
\be 
\left. \frac{q}{p}\right| _{B_s}=
\sqrt{\frac{(M_{12}^{B_s})^*}{M^{B_s}_{12}}}(1-\frac{r_{B_s}}{2}
\sin\zeta _{B_s})
\ee
and 
\be
\Delta M_{B_s}\simeq -2\overline M^{B_s}_{12}~~~~~\mbox{and}~~~~~
\Delta \Gamma_{B_s}\simeq 2\overline \Gamma_{12}^{B_s} 
\ee
where we have used the prediction that $\zeta_{B_s}$, unlike $\zeta_{B_d}$, 
is very small, since both $M_{12}$ and $\Gamma_{12}$ 
are dominated by contributions from the third and second family only.

%%%%%%%%%%%%%%%
\subsection{Evaluating the sign of $\Delta M_K$, 
$\Delta M_B$, and $\Delta\Gamma_B$ 
\label{SIGNM}}
%%%%%%%%%%%%
After having established a notation equally convenient 
for the $K$ and $B$ (as well as $D$) case we calculate 
$\Delta M_K$ and $\Delta M_B$. As is well known the 
dominant {\em short-distance} contribution to the 
effective $\Delta F=2$ interaction within the Standard 
Model is obtained from the box diagram. One finds 
\ba
&&{\cal H}_{eff}^{box}(\Delta F = 2, \mu ) = 
\left( \frac{G_F}{4\pi}\right) ^2 M_W^2 \cdot\nn 
&\cdot& \left[\eta _{cc}(\mu ) \lambda _c^2 E(x_c)  + 
\eta _{tt}(\mu ) \lambda _t^2 E(x_t)  + 
2\eta _{ct}(\mu ) \lambda _c \lambda _t E(x_c,x_t)  \right] 
[\overline q\gamma_\mu(1-\gamma_5)Q]^2 + h.c.
\mlab{SBOX}
\ea
where $Q=s,b$, and $q=d,s$; 
$\lambda ^Q_i$ denote combinations of KM parameters 
\be 
\lambda _i^Q = {\bf V}_{iQ}{\bf V}^*_{iq} \; , \; i=c,t \; , 
\ee 
and $E(x_i)$ and $E(x_i,x_j)$ reflect the box loops with equal 
and different internal quarks (charm or top), respectively: 
\be 
E(x_i) =  x_i \left( \frac{1}{4} + \frac{9}{4(1-x_i)} 
- \frac{3}{2(1-x_i)^2} \right) - 
\frac{3}{2} \left( \frac{x_i}{1- x_i}\right) ^3 {\rm log} x_i 
\ee
$$
E(x_c, x_t) = x_c x_t \left[ \left( \frac{1}{4} +   
\frac{3}{2}\frac{1}{1-x_t} - 
\frac{3}{4} \frac{1}{(1- x_t)^2} \right)    
\frac{{\rm log}x_t}{x_t - x_c} + (x_c \leftrightarrow  x_t) 
- \right. 
$$
\be 
\left. - \frac{3}{4} \frac{1}{(1-x_c)(1-x_t)}\right] \; ; \; 
x_i = \frac{m_i^2}{M_W^2} \; . 
\mlab{LIM}
\ee 
The $\eta _{qq'}$ contain the QCD radiative corrections; 
they have been studied through next-to-leading 
level in order 
to understand
the theoretical errors. We shall not go into the scale 
dependence 
as well as
errors associated with uncertainties in $\Lambda_{QCD}$, 
$m_t$, etc. 
Such a discussion can be found in Ref.\cite{buras}. For us 
it is 
important to note that they are all {\em positive}. 

These QCD corrections arise from evolving the effective 
Lagrangian from $M_W$ down to the scale $\mu$ at which 
the hadronic expectation value is evaluated. 
The latter task is far from trivial even for 
a local four-fermion operator since 
on-shell matrix elements are controlled by long-distance 
dynamics.  
{\em In principle} the value of $\mu$ does not matter: 
the $\mu$ dependance of the $\Delta F=2$ operator is 
compensated for by the $\mu$ dependance of the 
expectation value. {\em In practise} however, the 
available methods for calculating these matrix 
elements do not allow us to reliably track their 
$\mu$ dependance or they are applicable only for 
$\mu \sim 1$ GeV. 
Its size is customarily expressed as follows: 
\be 
\matel{P}{(\overline q  \gamma _{\mu}(1-\gamma_5)Q)
(\overline q  \gamma _{\mu}(1-\gamma_5)Q)}{\overline P} = 
-\frac{4}{3} B_P F_P^2 M_P \; , \bar P = [Q\bar q] 
\mlab{MES} 
\ee 
which represents a parametrization rather than an ansatz 
as long 
as the value of $B_P$ is left open. 

$B_P =1$ is referred to as {\em vacuum saturation} 
(VS) since it 
emerges when only the vacuum is inserted as intermediate 
state. Let us see the origin of the minus sign. One obtains   
\be 
\left. \matel{P}{(\overline q  \gamma _{\mu}(1-\gamma_5)Q)
(\overline q  \gamma _{\mu}(1-\gamma_5)Q)}{\overline P}
\right| _{VS} = \frac{4}{3} 
\matel{P}{(\overline q  \gamma _{\mu}(1-\gamma_5)Q)}{0} 
\matel{0}{
(\overline q  \gamma _{\mu}(1-\gamma_5)Q)}{\overline P}
\ee 
upon inserting the vacuum intermediate state in the two types 
of box diagrams; this yields the colour factor 
$1 + 1/N_C = \frac{4}{3}$. Setting
\be
\matel{0}{
(\overline q  \gamma _{\mu}(1-\gamma_5)Q)}{\overline P(k)}= 
iF_P k_{\mu} 
\ee
one finds with $\cp~|0\rangle = |0\rangle$ and the convention $\cp|P\ket=+|\overline P\ket$\footnote{The VS result follows even with 
$\cp~|0\rangle = e^{i\alpha}|0\rangle$.}

%%%%%%%%%%%%%%%%%%%%%%%%%%%%%%%%%%%%%%%%%%%
$$ 
\matel{P}{(\overline q  \gamma _{\mu}(1-\gamma_5)Q)}{0} = 
\matel{0}{(\overline Q  \gamma _{\mu}(1-\gamma_5)q)}
{P}^{\dagger} = 
$$ 
\be 
(\matel{0}{\cp ^{\dagger}\cp
(\overline Q  \gamma _{\mu}(1-\gamma_5)q)\cp ^{\dagger}\cp}
{P})^{\dagger}= 
- (\matel{0}{
(\overline q  \gamma _{\mu}(1-\gamma_5)Q)}
{\bar P})^{\dagger}= iF_P k_{\mu} 
\ee 
since 
\be
\cp\overline Q(\vec x)\gamma^\mu\gamma_5 q(\vec x)
\cp^
\dagger=
-\overline q(-\vec x)\gamma^\mu\gamma_5 Q(-\vec x).
\ee   

Several theoretical techniques have been employed to 
estimate the size of $B_P$. For a recent review, see Ref.
\cite{sonirev}.
Again for us it is important that they all yield $B_P>0$. 
\footnote{The fudge factor $B_P$ is sometimes called 
the bag factor since the MIT bag model was at an earlier time 
considered to yield a relatively reliable estimate for its 
size. It might be amusing to note while the bag model yields 
$B_B > 0$ as a very robust result, it is much less firm on 
$B_K$: both $B_K > 0$ as well as $B_K < 0$ emerge when varying 
the model parameters over a reasonable range!}
Within the Standard Model the short-distance contributions 
thus predict unequivocally 
\be 
M_2 - M_1 > 0 
\ee 
for $K$ as well as $B$ mesons. 
There are sizeable long distance contributions to 
$\Delta M_K$, which are estimated to be positive as well 
\cite{OURPAPER}. In any case it would be quite contrived to 
expect them to cancel the short distance contributions.

With the definition $\cp|P\ket = |\overline P\ket$
used here, $K_2$ is the mainly \cp~odd state.
Thus the Standard Model agrees with the experimental finding that 
$K_L$ is heavier than $K_S$ - a fact which is not often stressed 
in the literature.

For the K meson system, the box diagram does not lead to  
a reliable prediction for $\Delta\Gamma_K$ since the decay is dominated by 
$K\to\pi\pi$ which is anything but short distance dominated. 
The situation is quite
different for $\Delta \Gamma_{B_s}$ which is driven by
$[b\bar s]\to"c\bar c"\to[s\bar b]$ transitions. 
\footnote{It is quite amusing that for $\Delta\Gamma_K$ the $s\bar d\to"u\bar u"\to d\bar s$ transition gives the correct sign for $\Delta\Gamma_K$.}
It is reasonable to expect
a short distance treatment to provide at least 
a semi-quantitative description. Hence we predict\footnote{The question of the sign of $\Delta\Gamma_{B_d}$ might remain academic for experimental reasons. Theoretically the situation is not so clear due to large GIM cancellations\cite{BKUS}.}

\be
\Delta\Gamma_{B_s}=\Gamma_{B_s,even}-\Gamma_{B_s,odd}>0.
\ee
%%%%%%%%%%%%
\section{The phase of $\frac{q}{p}\overline\rho(\psi K_S)$
\label{PHASE}}
%%%%%%%%%%%%%
Now that we know $\sin(\Delta Mt)$ describes the oscillations  
with positive 
$\Delta M$, we need to know the phase of 
$\overline\rho(\psi K_S)$, which is 
defined by
\be
\overline\rho(\psi K_S)=\frac{\mat{\psi K_S}{H_{\Delta B=1}}
{\overline B}}
{\mat{\psi K_S}{H_{\Delta B=-1}}{B}}=\frac{\bra \psi K_S|\psi 
\overline K^0\ket}{\bra \psi K_S|\psi K^0\ket}
\frac{\mat{\psi \overline K^0}{H_{\Delta B=1}}{\overline B}}
{\mat{\psi K^0}{H_{\Delta B=-1}}{B}} \; ; 
\ee
here we have used 
$\cp H_{\Delta B=1}\cp^\dagger=H_{\Delta B=-1}^*$.

Adopting the convention 
$\cp~|P\rangle = + |\bar P \rangle $
we obtain
\ba
\mat{\psi \overline K^0}{H_{\Delta B=1}}{\overline B_d}&=&
\mat{\psi \overline K^0}{\cp^\dagger\cp H_{\Delta B=1}
\cp^\dagger\cp}
{\overline B_d}\nn
&=&-\mat{\psi K^0}{H_{\Delta B=-1}^*}{B_d}
\ea
where the minus sign reflects the fact that $\psi K^0$ 
form a P wave. Thus 
\be
\overline\rho(\psi K_S)=
-\frac{\mat{\psi K_S}{H_{\Delta B=-1}^*}{B_d}}
{\mat{\psi K_S}{H_{\Delta B=-1}}{B_d}}=-
\frac{{\bf V}_{cd}^*{\bf V}_{cs}}{{\bf V}_{cd}{\bf V}_{cs}^*}
\frac{{\bf V}_{cb}{\bf V}_{cs}^*}{{\bf V}_{cb}^*{\bf V}_{cs}}
\ee
where
$V^*_{cd}V_{cs}/V_{cd}V^*_{cs}\sim \left(\frac{q_K}{p_K}\right)^*$.
From \mref{qoverp}, we see that
\be
\frac{q}{p}=\sqrt{\frac{M_{12}^*}{M_{12}}}+{\cal O}(r)\propto
\frac{{\bf V}_{tb}^*{\bf V}_{td}}{{\bf V}_{tb}{\bf V}_{td}^*}
\ee
Putting everything together we find the asymmetry is given by
$$ 
{\rm sin} \Delta M_Bt \cdot 
\Im\left(\frac{q}{p}\overline\rho(\psi K_S)\right)=
- {\rm sin} |\Delta M_Bt| \cdot
\Im\left(\frac{{\bf V}_{tb}^*{\bf V}_{td}}
{{\bf V}_{tb}{\bf V}_{td}^*}
\frac{{\bf V}^*_{cd}}{{\bf V}_{cd}}
\frac{{\bf V}_{cb}}{{\bf V}_{cb}^*}\right) \simeq 
$$ 

\be 
\simeq  {\rm sin} |\Delta M_Bt| \cdot 
\frac{2 \eta (1-\rho )}{(1- \rho )^2 + \eta ^2} \; , 
\ee
with the quantities $\eta$ and $\rho$ referring to the 
Wolfenstein representation of the CKM matrix. 

With $\eta$ inferred to be positive from $\epsilon$ one 
concludes that this 
asymmetry has to be {\em positive}! 

%%%%%%%%%%%%
\section{Different conventions \label{OT}}
%%%%%%%%%%%%
%%%%%%%%%%%%%%%%%%%%
\subsection{Using $\cp~|P\rangle = - |\bar P\rangle$}
%%%%%%%%%%%%%%%%
It is instructive to analyze what happens if one 
uses the equally allowed convention 
\be 
\cp~|P\rangle = - |\bar P\rangle \; . 
\ee 
Various intermediate expressions change; e.g., 
repeating the steps of Sect. \ref{SIGNM} one has 
\be  
\left. 
\matel{P}{(\overline q  \gamma _{\mu}(1-\gamma_5)Q)
(\overline d  \gamma _{\mu}(1-\gamma_5)s)}{\overline P}
\right| _{VS} = 
+\frac{4}{3} F_P^2 M_P \; , \bar P = [Q\bar q] 
\mlab{MESNEG} 
\ee 
since now  
\be 
\matel{P}{(\overline q  \gamma _{\mu}(1-\gamma_5)Q)}{0} = 
-iF_P k_{\mu} 
\ee 
That means that $\bar M_{12}$ changes sign; yet -- as pointed out 
before -- the two labels $1$ and $2$ exchange their roles then 
as well, see \mref{B3}. The prediction that $K_L$ is slightly heavier than 
$K_S$ thus remains unaffected. 

Furthermore the sign of $\bar \rho (\psi K_S)$ changes 
as well, see Sect. \ref{PHASE}. Thus the combination 
sin$\Delta M_Bt \, \cdot$ 
Im$\frac{q}{p}\bar \rho (\psi K_S)$ remains unchanged!  

The ambiguity we have in the sign of $\Delta M_B$ is 
thus compensated by a corresponding ambiguity 
in the sign of $\frac{q}{p}\overline\rho(\psi K_S)$.
We can see this also in the following way:
\begin{enumerate}
\item
Changing $q/p \to -q/p$ maintains the defining property 
$(q/p)^2=(M_{12}^{*}-\frac{i}{2}\Gamma^*_{12})/(M_{12}-\frac{i}{2}\Gamma_{12})$,
see \mref{qoverp}, and cannot affect observables.
\item
Yet the two mass eigenstates labeled by subscripts 1 and 2 exchange places, 
see \mref{P1P2gen}. 
\item
The difference $\Delta M= -2{\rm Re}[\frac{q}{p}
(M_{12}-\frac{i}{2}\Gamma_{12})]$ then flips its sign - yet 
so does ${\rm Im}\frac{q}{p}\overline\rho(\psi K_S)$!
\item
The product $(\sin \Delta M_Bt)\cdot{\rm Im}\frac{q}{p}\overline\rho(\psi K_S)$ therefore remains invariant. 
\end{enumerate}

%%%%%%%%%%%%%%
\subsection{Another approach}
%%%%%%%%%%%%%
The authors of \cite{qn} define 
\be
\Delta M_B=M_H-M_L
\ee
where H [L] stands for heavy [light]; $\Delta M$ is 
thus positive 
by definition.
They also define 
\ba
|B_L\ket&=&p|B^0\ket+q|\overline B^0\ket\nn
|B_H\ket&=&p|B^0\ket-q|\overline B^0\ket
\ea
For clearity, let us imagine a world in which \cp~is not violated.
Then it is sensible to ask if the \cp~ even state is heavier 
or lighter
than the \cp~odd state. For good reason they have not decided 
on this question,
since there is a freedom in the sign of
\be
\frac{q}{p}=\pm\sqrt{\frac{M_{12}^*-\frac{i}{2}\Gamma_{12}^*}
{M_{12}-\frac{i}{2}\Gamma_{12}}}
\mlab{poverq+-} 
\ee
If in computing the mass difference $M_2 - M_1$ turns out to be 
negative -- yet we want to keep the convention 
$\cp~|P\rangle = |\bar P \rangle$ -- we must choose the 
minus sign in \mref{poverq+-}. To say it differently: 
if we require $\Delta M_B >0$  we have to compute 
the sign of $M_2 - M_1$ to decide on the sign for 
$q/p$ in \mref{poverq+-}. Alternatively we can keep the 
plus sign if we adopt $\cp~|B^0\rangle = - |\bar B^0\rangle$. 
In either case we must compute the sign of $M_2-M_1$. 
The convention we must adopt is not fixed in this approach
until we
perform the theoretical computation of $M_2-M_1$.
We feel it is more natural to define a convention independent of prior
theoretical computations.
%%%%%%%%%
%%%%%%%%%%%%%%%%%%%%%%%%%%%%%%%%%%%%%%%%%%%%
\begin{figure}[t]
%\vspace{8cm}
\epsfxsize=10cm
\centerline{\epsfbox{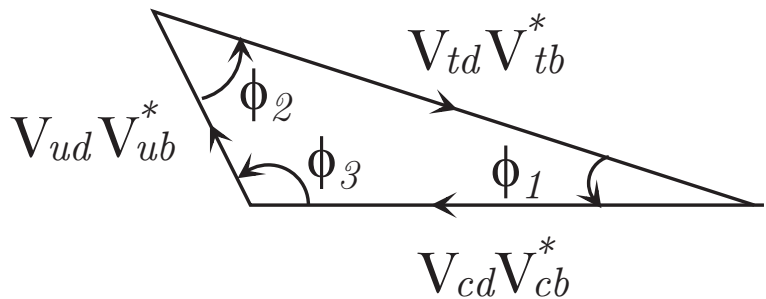}}
\caption{The unitarity triangle for B decays.}
\mlabf{unit2}
\end{figure}
%%%%%%%%%%%%%%%%%%%%%%%%%%%%%%%%%%%%%%%%%%%%%%%%%

%%%%%%%%%%%
\section{Relation to the unitarity triangle 
\label{TRIANGLES}}

To conclude our discussion we 
explicitely restate the connections between 
the angles in the unitarity triangle, the CKM parameters and 
observable \cp~asymmetries. 

The angles $\phi _1$, $\phi _2$ and $\phi _3$ are 
defined in \mreff{unit2}; as can be read off from \mreff{unit1} 
they can be expressed by   
\ba
\phi_1&=&\pi-\arg\left(\frac{-{\bf V}^*_{tb}{\bf V}_{td}}
{-{\bf V}^*_{cb}{\bf V}_{cd}}\right),\nn
\phi_2&=&\arg\left(\frac{{\bf V}^*_{tb}{\bf V}_{td}}
{-{\bf V}^*_{ub}{\bf V}_{ud}}\right),\nn
\phi_3&=&\arg\left(\frac{{\bf V}^*_{ub}{\bf V}_{ud}}
{-{\bf V}^*_{cb}{\bf V}_{cd}}\right).
\mlab{15.17}
\ea

The \cp~asymmetry in $B_d \to \psi K_S$, which is driven 
by a single $\Delta B =1$ operator, is given by 
$$ 
A_{\psi K_S} 
= {\rm sin}(\Delta M_Bt) 
\Im\left(\frac{q}{p}\overline\rho(\psi K_S)\right)= 
$$
\be 
= \sin(2\phi_1)\sin(|\Delta M_B|t)
\ee 
To the degree we can eliminate the Penguin contribution 
to $B_d \to \pi ^+ \pi ^-$ by, say, extracting the 
asymmetry for the isopin-two $\pi \pi$ final state, we 
have likewise: 
$$  
A_{(\pi \pi )_{I=2}} = {\rm sin}(\Delta M_Bt) 
\Im\left( \frac{q}{p}\overline\rho([\pi \pi]_{I=2})  
\right)= 
$$ 
\be
= \Im \left( \frac{V_{tb}V^*_{td}}{V^*_{tb}V_{td}} 
\frac{V_{ud}V^*_{ub}}{V^*_{ud}V_{ub}}\right) 
{\rm sin}(|\Delta M_Bt|) \simeq 
{\rm sin}(|\Delta M_Bt|) \sin (2\phi_2) 
\ee  

%%%%%%%%%%%%%%%%%%%%%%%%%%%%%%%%%%%%%%%%%%%%
\begin{figure}[t]
%\vspace{8cm}
\epsfxsize=15cm
\centerline{\epsfbox{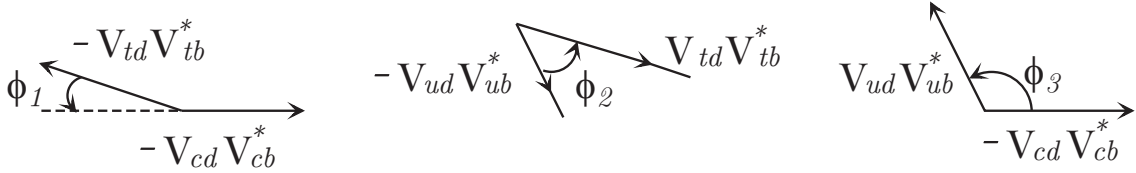}}
\caption{The unitarity triangle.}
\mlabf{unit1}
\end{figure}
%%%%%%%%%%%%%%%%%%%%%%%%%%%%%%%%%%%%%%%%%%%%%%%%%

%%%%%%%%%%%%%
\section{Summary \label{SUMMARY}} 
%%%%%%%%%%

We have shown that the {\em sign} of the \cp~asymmetry in 
$B_d \to \psi K_S$ can be predicted within a {\em given 
theory} for $\Delta M_B$ and $\frac{q}{p}\bar \rho (\psi K_S)$ 
even if the sign of $\Delta M_B$ cannot be 
determined experimentally. 
The same holds for  
$B_d \to \pi \pi$ etc. if one knows the weak operators 
driving these transitions. 

En passant we have reminded the reader that the Standard 
Model can reproduce the observed sign of 
$\Delta M_K$ (and roughly its magnitude) through 
{\em short-distance} dynamics 
\footnote{In the charm complex on the other hand 
$\Delta M_D$ is dominated by long distance dynamics within 
the Standard Model.}. 

Our discussion illustrates that we can choose any 
convention we want -- provided care is applied in treating 
everything consistently. We view it, however, as very useful 
to have a formalism for describing $B$ oscillations 
that parallels that for kaons. Furthermore it is 
much more natural to invoke theory to decide on the sign 
of $\Delta M_B$.  

\sk
\centerline{Acknowledgements}
Work of IIB has been supported in part by the NSF under the grant number PHY 96-0508. Work of AIS has been supported in part by Grant-in-Aid for Special Project Research (Physics of \cp~violation).

%%%%%%%%%%%%%%%%%%%%%%%%%%%%%%%

\end{document}

%% file: sanda.tex
% equations
\def\bfix{~\newline\centerline{XXXXXXXXXXX corrected versionXXXXXXXXXXXXXX}\newline}
\def\efix{~\newline\centerline{XXXXXXXXXXXX corrected version ends hereXXXX}\newline}
\def\bXX{~\newline \centerline{XXXXXXXXXXXXXX this part is correctedXXXXXXX}\newline}
\def\eXX{~\newline \centerline{XXXXXXXXXXXXXXXXXXXXXXXXXXXXXXXXXXXXX}\newline}
\def\ket{{\rangle}}
\def\bra{{\langle}}
\def\ie{{\it i.e.}}
\def\be{\begin{equation}}
\def\ee{\end{equation}}
\def\ba{\begin{eqnarray}}
\def\ea{\end{eqnarray}}
\def\mref#1{Eq.(\ref{Eq:#1})}
\def\mreff#1{Fig.\ref{Eq:#1}}
\def\mreft#1{Table\ref{Eq:#1}}
\def\mlab#1{\label{Eq:#1}}
\def\mlabf#1{\label{Eq:#1}}
\def\mlabt#1{\label{Eq:#1}}
\def\half{\frac{1}{2}}
\def\to{\rightarrow}
\def\nn{\nonumber\\}
\def\sk{\vskip 1cm}
\def\skk{\vskip 3mm}
\def\mat#1#2#3{\langle{#1}\vert{#2}\vert{#3}\rangle}
\def\etal{{\it et al.}}
\def\etc{{\it etc.~}}
\def\eg{{\it e.g.~}}

%%%%%%%%%%%%%%%%%%%%%%%%%%%%%%%%%
\def\ibid#1#2#3{{\it ibid.}{\bf #1} #2 {(#3)} }
\def\PR#1#2#3 {{\it Phys. Rev. }{\bf D#1} #2 {(#3)} }
\def\PRL#1#2#3 {{\it Phys. Rev. Lett. }{\bf #1} #2 {(#3)} }
\def\PL#1#2#3 {{\it Phys. Lett. }{\bf #1} #2 {(#3)}  }
\def\AP#1#2#3 {{\it Ann, Phys. }{\bf #1} #2 {(#3)} }
\def\ZP#1#2#3 {{\it Z. Phys. }{\bf #1} #2 {(#3)} }
\def\NP#1#2#3 {{\it Nucl. Phys. }{\bf #1} #2 {(#3)}  }
\def\MPL#1#2#3 {{\it Mod. Phys. Lett.}{\bf #1} #2 {(#3)}  }
\def\NC#1#2#3 {{\it Nuov. Cimm. }{\bf #1} #2 {(#3)}  }
\def\PREP#1#2#3 {{\it Phys. Report }{\bf #1} #2 {(#3)}  }
\def\PROG#1#2#3 {{\it Prog. Theor. Phys. }{\bf #1} #2 {(#3)}   }
\def\SOV#1#2#3{{\it Sov. J. Nucl. Phys. }{\bf #1} #2 {(#3)}   }
\def\JETP#1#2#3{{\it JETP}{\bf #1} #2 {(#3)}   }
\def\RMP#1#2#3{{\it Rev. Mod. Phys.}{\bf #1} #2 {(#3)}   }
%matrix element

%%%%%%%%%%%%%%%%%%%%%%%%%%%%%%%%%%%%%%%%%%%%%%%%%%%%%%%%%%%%%%%%%%%%%
\def\phiout{{\phi_a^{out}}}
\def\phiin{\phi_a^{in}}
\def\psiout{\psi_a^{out}}
\def\psiin{\psi_a^{in}}
\def\vin{\phi_{a\mu}^{in}}
\def\vout{\phi_{a\mu}^{out}}
\def\ofx{{(x)}}
\def\op{{\bf P}}
\def\oc{{\bf C}}
\def\ot{{\bf T}}
\def\cp{{\bf CP}}
\def\cpt{{\bf CPT}}
\def\vecr{{\vec r}}
\def\vecn{{\vec {\nabla}}}
\def\vecA{{\vec A}}
\def\psibar{\overline\psi}
\def\outin{{{out}\choose{in}}}
\def\inout{{{in}\choose{out}}}

\def\vr{\vec x}
\def\itt{{\it T}}
\def\b{{\bf b}}
\def\a{{\bf a}}
\def\d{{\bf d}}

\def\alphadot{{\dot\alpha}}
\def\betadot{{\dot\beta}}
\def\gammadot{{\dot\gamma}}

\def\N{\sqrt{{{E+mc^2}\over{2mc^2}}}}
\def\F#1{{{#1}\over{E+mc^2}}}
\def\itemm{\hangindent\parindent\textindent}
\def\noi{\noindent}
\def\onehead#1{\vskip1pc\leftline{\bf #1}}
\def\twohead#1{\vskip1pc\leftline{\bf #1}}
\def\ts{\thinspace}

\def\sq2{{1\over{\sqrt{2}}}}
\def\omegaar{{\vec{\omega}}}
\def\kbar{\overline K}
\def\Pbar{\overline P}
\def\Abar{\overline A}
\def\dbar{\overline d}
\def\ubar{\overline u}
\def\sbar{\overline s}
\def\bbar{\overline B}
\def\gbar{\overline g}
\def\pbar{\overline P}
\def\qbar{\overline q}
\def\vecr{\vec r}

\def\dm{\Delta m}
\def\ss{(1+s^2)}
\def\cdmp{cos\Delta m(t_1+t_2)}
\def\cdm{cos\Delta m(t_1-t_2)}

\def\g5{\gamma_5}
\def\gm{(1-\gamma_5)}
\def\gp{(1+\gamma_5)}

% vectors
\def\mvec#1{\vec{#1}\,}
\def\pslash{\mbox{/\llap p}}
\def\slash#1{\mbox{/\llap #1}}

% small roman indices
\def\msmall#1{\mbox{\rm \small #1}}
%%%%%%%%%%%%%%%%%%%%%%%%%%%%%%%%%%%%%%%%%%%%%%%%%
%bigi
%\renewcommand{\topfraction}{1.0}
%\renewcommand{\bottomfraction}{1.0}
%\renewcommand{\textfraction}{0.0}
\newcommand{\matel}[3]{\langle #1|#2|#3\rangle}
\newcommand{\hscale}{\mu\ind{hadr}}
\newcommand{\aver}[1]{\langle #1\rangle} 
\renewcommand{\Im}{\mbox{Im}\,}
\renewcommand{\Re}{\mbox{Re}\,}
\newcommand{\GeV}{\,\mbox{GeV}}
\newcommand{\MeV}{\,\mbox{MeV}}
\newcommand{\BR}{\,\mbox{BR}}